# Compact Differential Phase-Shift Quantum Receiver Assisted by a SOI / BiCMOS Micro-Ring Resonator

**Nemanja Vokić, Dinka Milovančev, Winfried Boxleitner, Hannes Hübel, and Bernhard Schrenk**
*AIT Austrian Institute of Technology, Center for Digital Safety&Security / Security & Communication Technologies, 1210 Vienna, Austria.*
*Author e-mail address: bernhard.schrenk@ait.ac.at*

We demonstrate a phase-selective and colorless quantum receiver assisted by a silicon-on-insulator microring, enabling a low 1.3% QBER at 5.3kb/s secure-key rate. No penalty incurs compared to a delay interferometer. BiCMOS 3D-integration is proven feasible.

## 1. Introduction

Quantum key distribution (QKD) builds on the fundamental laws of quantum mechanics to establish a secret key between two communication terminals and to identify a potential eavesdropper. A wide range of QKD protocols have been experimentally implemented and partially commercialized. However, the cost-effectiveness of QKD systems is greatly limited by the applied component technologies for generating and receiving quantum states, leading to rack systems with costs incurring several tens of k$. We have recently proposed a network-level approach for differential phase shift (DPS) QKD, which allows to off-load complexity associated to the interferometric quantum receiver at a centralized location [1,2]. Although this allows cost-sharing of rather expensive components, this approach cannot be adopted by point-to-point links as found in datacenter architectures. To support an economic QKD model in this short-reach realm, further simplification in terms of applied receiver technology is required.

In this work we demonstrate, for the first time to our best knowledge, an interferometer-less receiver architecture for DPS-QKD. A low form-factor is obtained by substituting the typically used 1-GHz Mach-Zehnder interferometer (MZI) by a compact micro-ring resonator (MRR) realized on the silicon-on-insulator platform. Experimental measurements demonstrate penalty-free operation at a low quantum bit error ratio (QBER) of 1.9% with the MRR-based receiver. We further transfer the MRR to a BiCMOS wafer and elaborate on the incurred performance penalty.

## 2. Differential Phase Shift QKD with MRR-Assisted Receiver

DPS QKD is a main contender to realize simplistic quantum links (Fig. 1). It encodes information in the optical phase difference between two successive pulses through either active opto-electronic modulation [1] or passive path choice in an unbalanced MZI [3]. At the receiver, another MZI typically acts as delay interferometer before single-photon detection of the quantum signal, whose rather low 1-GHz line rate leads to bulky MZI implementations. Although the DPS scheme features a simplistic reception scheme in only one measurement basis and uses the full rate of received photons in contrary to other QKD protocols that require a measurement in two non-orthogonal settings and digital sifting, its interferometric elements require careful alignment to the optical carrier wavelength. As one approach to minimize the number of MZIs, a chirp-modulated laser can greatly simplify the DPS transmitter by substituting its MZI or an active LiNbO3 phase modulator [2], to ultimately yield a quantum transmitter that actually adheres to the SFP form-factor. In this work, we tackle the remaining challenge of alleviating the receiver from its MZI by replacing it with a much smaller silicon MRR (Fig. 1). Such resonators have been evaluated in combination with the demodulation of classical DPSK signals [4] and offer a periodic transfer function with high Q-factor at small footprint, as sought for operation at low symbol rates in a wide wavelength range for the data signal.

The resonator used in this work is a MRR that was fabricated on silicon-on-insulator (SOI) technology using a 220×500 nm$^2$ waveguide cross-section. The diameter of the MRR is 200 µm. Figure 2 plots the measured fiber-to-fiber transfer function of the through-port, as it is used for DPS demodulation. The fiber-to-fiber coupling is -16.7 dB. The FWHM bandwidth of the spectral notch at the through-port is 0.27 GHz and the free spectral range is 120.1 GHz. The peak extinction is 23.7 dB and reduces to 17.5 dB at an offset of 1 GHz from the dip. It shall be noted that a single-polarization MRR is used in this work; however, polarization-immune operation can be in principle supported by SOI platforms.

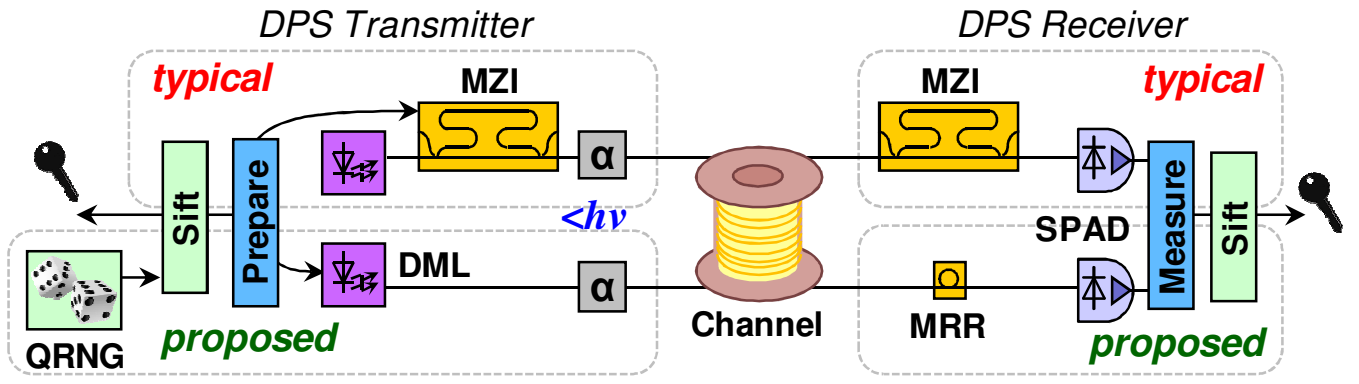

Fig. 1. Typical and proposed DPS system after component simplification.

In order to obtain a low QBER close to the intrinsic QBER for quantum signal reception, a high extinction is important for the MRR. The single-photon avalanche photodetector (SPAD) after the MRR demodulator acts as hard-limiter with a low detection efficiency of η ~ 10% and a constant dark count rate δ ~ 550 c/s. Space bits cannot be used for the purpose of error counting as these cannot be distinguished from absorbed photons in the quantum signal, which is launched from the transmitter at a symbol rate of $R_{sym}$ = 1 GHz and an average photon number of μ = 0.1 photons/symbol. By ensuring a high visibility for the optical phase-to-amplitude demodulation, provided through a high extinction $\varepsilon$ of the MRR, the mark bits are correctly registered. Moreover, SPADs are suffering from after-pulsing. Together with the optical loss $L$ between transmitter and SPAD, hence including the optical demodulator, the QBER can be estimated as ratio between erroneous bits deriving from erroneous phase demodulation and detector imperfection, to the total number of received bits. Figure 2c discusses this analytic QBER dependence on $\varepsilon$ for the aforementioned parameters and a loss budget of $L$ = 23.5 dB, as it is found as optimum value between excessive dark counts and after-pulsing due to SPAD saturation. The QBER of 1.93% for $\varepsilon$ = 18 dB gives a large margin towards the threshold of 5% for secret-key generation in DPS QKD [5]. Experimental results for three different MRR samples, as will be presented shortly, have been included in Fig. 2c. There is good agreement with the model, with an deviation of 0.3% at $\varepsilon$ = 18 dB.

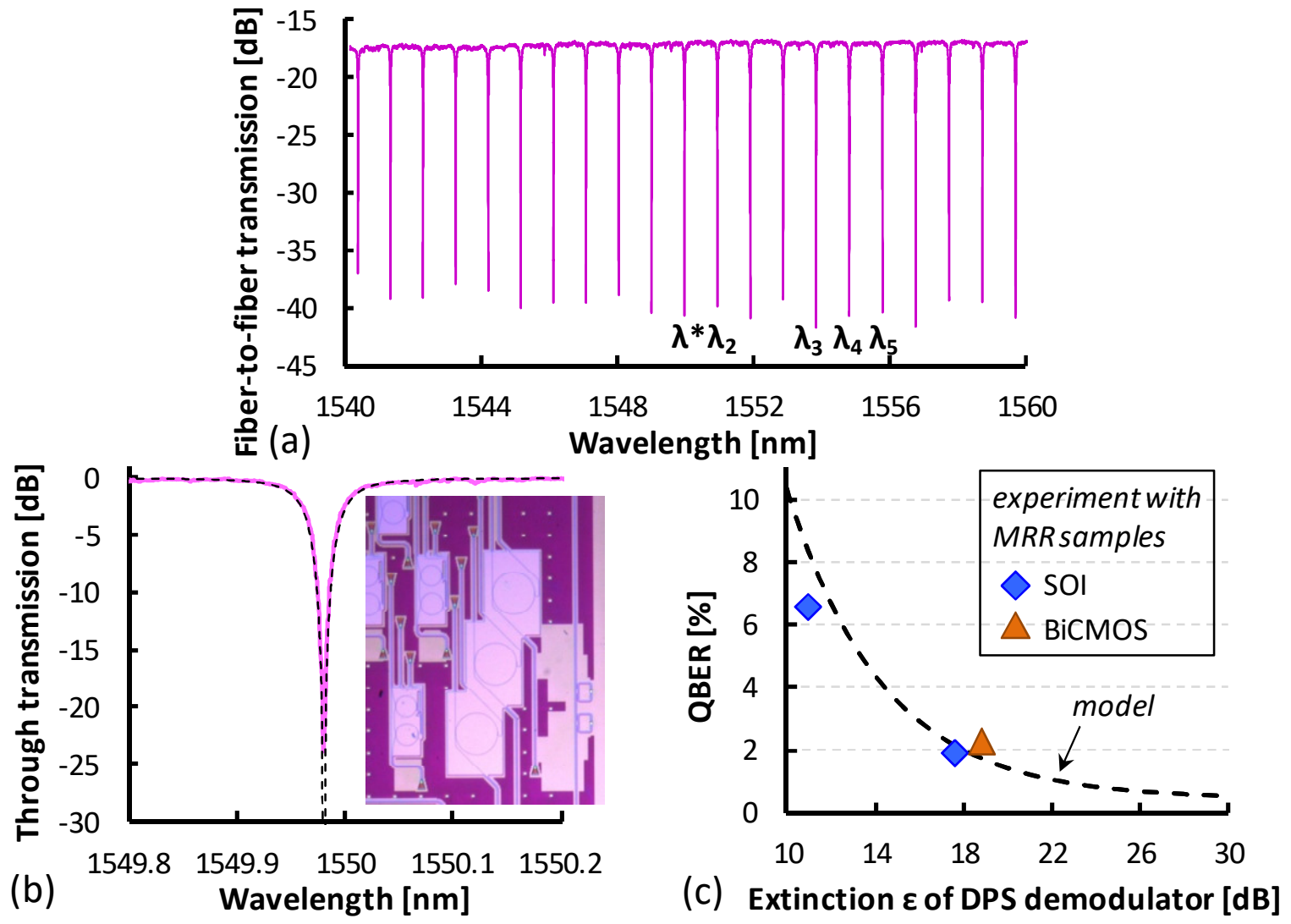

Fig. 2. (a) MRR fiber-to-fiber transfer function of through port and (b) close-up of a notch. (c) QBER as function of demodulator extinction.

## 3. Experimental DPS-QKD Setup with MRR- and MZI-Based Receiver

The experimental setup used to evaluate the MRR-based DPS receiver in comparison to a standard MZI-based receiver is presented in Fig. 3a. The optical transmitter employs a DFB laser and a $LiNbO_3$ phase modulator (PM), which encodes DPS at 1 Gbaud using $\pi$-phase shift keying. Although a directly modulated laser [2] could simplify this arrangement, the present work seeks to investigate low QBER penalties at the lowest possible (i.e. intrinsic) QBER that is solely yielded by the dark counts. For this reason, any extra penalty that arises from a transmitter-side complexity-performance trade-off is to be avoided. After the transmission channel, which is implemented through a 25.7 km standard single-mode fiber (SMF) link, the DPS signal is demodulated by either the MRR (R in Fig. 3) or a 1-GHz silica-on-silicon MZI with a polarisation-dependent extinction of 16.6 to 24.7 dB (M). Manual polarization control (PC) ensures high coupling efficiency for the single-polarization MRR design and high extinction for the MZI. Polarization-immune operation would require to implement a diversity scheme for the MRR. A signal monitor comprised of a SOA+PIN with a 200 GHz bandpass filter has been included after the optical demodulator in order to evaluate the signal integrity and to perform classical BER measurements for comparison. Since the experimental arrangement induces loss between the optical emitter and the receiver, an EDFA was used at the transmitter. In a practical DPS-QKD arrangement without classical monitor after the demodulator, this optical amplifier would be omitted and replaced by a high optical attenuation so to ensure a low signal launch in the order of 0.1 photons/symbol. This attenuator ($A_{tt}$) has been interchangeably placed after the demodulator in the present work, before the DPS receiver. The latter includes a WDM filtering device to investigate the transmission performance at various wavelengths ($\lambda^*$, $\lambda_2$ …$\lambda_5$), and a free-running InGaAs SPAD with a detection efficiency of 10%. A time-tagging module (TTM) registers the detector events and allows for real-time estimation of raw key rate and QBER.

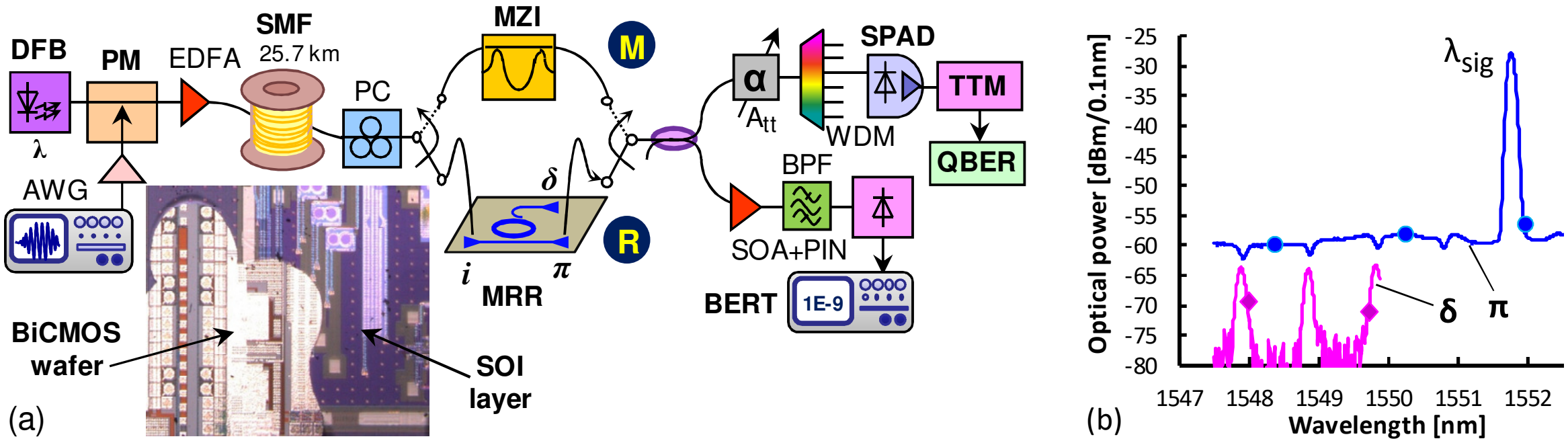

Fig. 3. (a) Experimental setup. (b) Spectral allocation of DPS signal at $\lambda_{sig}$ to the ring transfer functions of through (π) and drop (δ) ports.

## 4. Results and Discussion

Figure 3b shows the optical spectrum at the MRR through (π, ●) and drop (δ, ◆) port, with the DPS signal at $\lambda_{sig}$ being present at one of the MRR resonances, yielding the demodulated DPS data at the through-port. The performance of the MRR as DPS demodulator has been evaluated through classical BER and QBER measurements.

A clearly open eye diagram is acquired through the monitor receiver. A low back-to-back BER of $10^{-10}$ is obtained for a received power of -28.6 dBm at the SOA+PIN receiver (Fig. 4a). The high MRR extinction leads to a similar performance as obtained for the MZI-based demodulator, which features a 0.3 dB better sensitivity.

The back-to-back quantum reception performance was evaluated in terms of raw key rate (■) and QBER (▲) as function of the optical budget between the DPS transmitter, emitting at the quantum level (μ = 0.1 photons/symbol), and the SPAD (Fig. 4b). In this way SPAD-related effects such as after-pulsing and dark counts can be investigated. Temporal filtering within 30% of the symbol period is applied after detector event registration in order to suppress dark counts between the DPS symbols. A QBER minimum of 1.31% is obtained for a loss budget of 21.6 dB. This QBER is well below the 5% error threshold of DPS operation. A raw-key rate of 10.05 kb/s is obtained at this budget, with an estimated secret-key at 5.31 kb/s to remain after error correction and privacy amplification [5].

In order to prove the stability of the proposed demodulation scheme, QBER and rate measurements have been conducted over time (Fig. 4c). A back-to-back QBER of 1.93% and a raw key rate of 5.38 kb/s are obtained for a loss budget of 26.6 dB (point $\rho$ in Fig. 4b). There is no strong deviation, which indicates stable operation despite the high extinction of the MRR. A similar QBER value is obtained when re-allocating the transmitted DPS signal at a different WDM channel. The peak-to-peak deviation in QBER within the range from $\lambda^*$ = 1549.98 nm to $\lambda_5$ = 1555.77 nm was 0.59%. The DPS reception performance has been further evaluated for transmission over 25.7 km. In order to enable a comparison to the back-to-back measurements, the delivered power to the SPAD receiver has been kept constant. There was no degradation observed due to the present transmission fiber, which is evidenced from the low QBER of 1.84%, for which a key rate of $2.48\times10^{-6}$ secure bits/symbol can be obtained. The drop in key rate is attributed to the continuous polarisation drift in the transmission fiber span and the associated SOI coupling loss. Given the aforementioned fiber-to-fiber insertion loss for the MRR, a link budget of 9.9 dB is compatible. Direct coupling of the SPAD to the MRR without intermediate fiber can increase the link budget to 18.3 dB.

The 3D integration of passive demodulation function with photodetection would allow the realization of a fully-integrated quantum receiver. The SOI wafer has further been face-to-face bonded to a BiCMOS wafer containing electrical circuitry (Fig. 3a) in order to assess the impact of this co-integration step on the demodulator performance. After SOI substrate removal, a thin photonics layer resides on the BiCMOS wafer and light coupling can be accomplished through the buried oxide layer [6]. The fiber-to-fiber insertion loss for the MRR increased by 1.1 dB, however, an extinction of 27.3 dB remains at the resonant wavelength. With these a low QBER of 2.83% has been obtained for the BiCMOS wafer. This proves the successful co-integration towards an all-silicon DPS receiver.

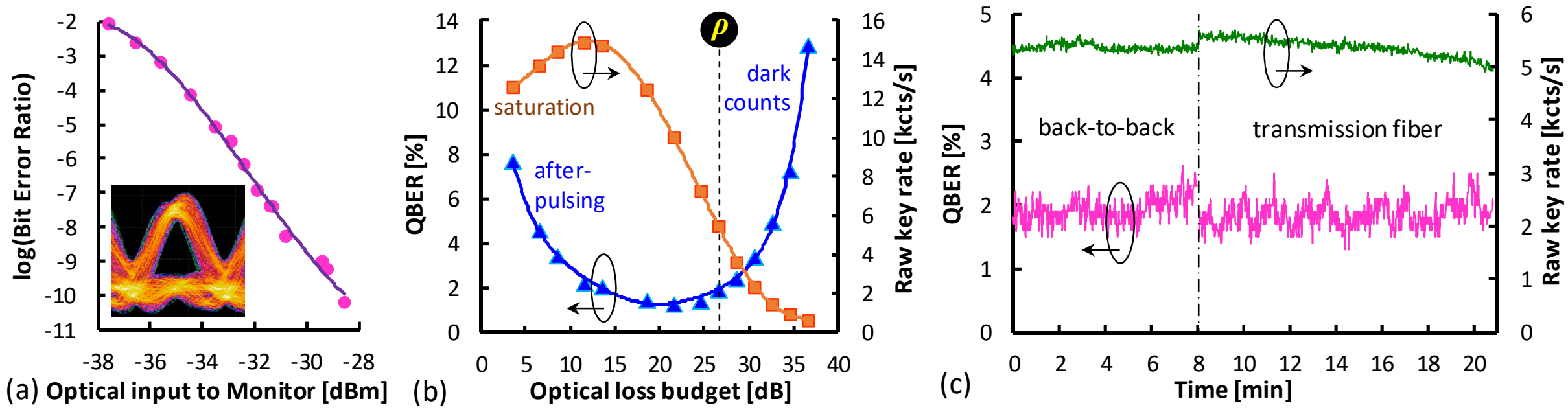

Fig. 4. (a) Classical BER after DPSK demodulation. (b) QBER and raw rate as function of delivered power. (c) Long-term DPS performance.

## 5. Conclusions

We have experimentally demonstrated the penalty-free substitution of a delay interferometer in a quantum receiver by a compact silicon MRR. Its high on-resonance extinction enables to demodulate DPS signals at a low QBER of 1.3%. Colorless DPS-QKD operation has been shown at a rate of $5.31\times10^{-6}$ secure bits/symbol for a loss budget of 26.6 dB. A similar performance is obtained after transferring the SOI ring to a BiCMOS wafer.

## 6. Acknowledgement

This work has received funding from the EU Horizon 2020 programme under grant agreement No 820474.